\documentclass[aps,prd,eqsecnum,showpacs,amsmath]{revtex4}
\usepackage{amsmath,amsfonts,amssymb,textcomp}
\usepackage{yfonts,bigints}
\usepackage{graphicx}
\newcommand{\gok}{\text{\textgoth{k}}}
\newcommand{\pd}{\partial}

\newcommand*{\No}{\textnumero}
\begin{document}

\title{Fab Four self-interaction in quantum regime}

\author{A.B. Arbuzov$^{1,2}$}
\author{B.N. Latosh$^{1,2}$}

\affiliation{
$^1$Bogoliubov Laboratory for Theoretical Physics, JINR, Dubna 141980, Russia, \\
$^2$ Dubna State University, Dubna 141982, Russia
}

\date{\today}

\begin{abstract}
  Quantum behavior of the John Lagrangian from the Fab Four class of Covariant Galileons is studied. We consider one-loop corrections to the John interaction due to cubic scalar field interactions. Counter Terms are calculated, one appears because of massless scalar field theory infared issues, another one lie in the George class, rest of them can be reduced to the initial lagrangian up to surface term. The role of quantum corrections in context of cosmological applications is discussed. 
\end{abstract}



\maketitle

\section{Introduction}

Scalar-tensor models of gravity is one of the simplest classes of modified gravity models, which dates back to the Brans-Dicke theory~\cite{Brans:1961sx}. Such models arise in different contexts and possess a wide phenomenology. For instance, they appear in the context of low-energy string models~\cite{Gross:1986mw}, models with auxiliary dimensions~\cite{Kim:1999dq}, inflation~\cite{Linde:1983gd} and wormholes~\cite{Roman:1992xj, Jamil:2009vn}. Some scalar-tensor models can pass the Solar system tests because of the screening mechanism~\cite{Vainshtein:1972sx,Babichev:2013usa} which makes them even more attractive from the theoretical point of view.

In recent times a particular class of scalar-tensor models called Galileons has drawn a lot of attention, see~\cite{Deffayet:2013lga} for a detailed review. This class was first discovered by Gregory Horndeski~\cite{Horndeski:1974wa} who searched for the most general scalar-tensor model of gravity with second order field equations. Later it was shown that the Horndeski theory is equivalent to the so-called Covariant Galileons~\cite{Kobayashi:2011nu}. Galileons originated from a low-dimensional reduction of braneworld models to a flat spacetime~\cite{Nicolis:2008in}. A compactification procedure spawns an additional scalar field with the so-called Generalized Galilean symmetry which protects the differential order of field equations. Covariant Galileons are the most simple generalization of the Galileons for a curved spacetime~\cite{Deffayet:2009wt,Deffayet:2009mn}. The generalization performed by the extension of the Galileons Lagrangian with curvature-related terms that cancel out higher derivatives thus preserving the order of field equations. This generalization comes with a price of the Generalized Galilean symmetry, but it is possible to construnct models that at least partially preserve the symmetry \cite{Pirtskhalava:2015nla, Germani:2011bc, Banerjee:2017qdl}. Covariant Galileons don't contain ghosts because of the order of field equations, but it is possible to construct even more general ghost-free scalar-tensor model of gravity with higher derivatives~\cite{Gleyzes:2014dya}. Such models are known as the beyond Horndeski ones and they can be reduced to the Galileons by a proper redefinition of dynamical variables.

In four dimensions the Covariant Galileons are generated by the following Lagrangians:
\begin{eqnarray}
  \mathcal{L}_{2} & = & G_{2} ,\\
  \mathcal{L}_{3} & = & - G_{3} \square \phi,\\
  \mathcal{L}_{4} & = & G_{4} R + G_{4X} \left[ (\square \phi)^{2} - (\nabla_{\mu} \nabla_{\nu} \phi)^{2} \right],\\
  \mathcal{L}_{5} & = & G_{5} G_{\mu\nu} \nabla^{\mu} \nabla^{\nu}\phi - \cfrac16 ~G_{5X} \left[ (\square \phi)^{3} -3 \square \phi (\nabla_{\mu} \nabla_{\nu} \phi)^{2} +2 (\nabla_{\mu} \nabla_{\nu} \phi)^{3} \right],
\end{eqnarray}
where $G_2$, $G_3$, $G_4$, and $G_5$ are arbitrary functions of the Galileon field $\phi$ and the standard kinetic term $X=1/2 ~\pd_\mu \phi \pd^\mu\phi$; subscript $X$ denotes the derivative with respect to the standard kinetic term; $R$ is the Ricci curvature, and $G_{\mu\nu}$ is the Einstein tensor.

The Covariant Galileons can be reduced to the General Relativity (GR) by setting $G_2=G_3=G_5=0$ and $G_4 = 1/16\pi G$. At the same time, the cosmological constant should be included in the model as a free parameter~\cite{Lovelock:1971yv} which is going to affect the cosmological properties of the Covariant Galileons. In ref.~\cite{Charmousis:2011bf} it was shown that there exists a narrow subclass of the Covariant Galileons which is able to screen the cosmological constant completely on the Friedmann-Lemaitre-Robertson-Walker background. This class is known as the Fab Four one, it is generated by the following Lagrangians:
\begin{eqnarray}
  L_\text{John} &=& V_J(\phi) G^{\mu\nu}\nabla_\mu\phi \nabla_\nu\phi ,\\
  L_\text{George} &=& V_G(\phi) R  ,\\
  L_\text{Ringo} &=& V_R(\phi)\hat G   ,\\
  L_\text{Paul} &=& V_P(\phi) P^{\mu\nu\alpha\beta} \nabla_\mu\phi \nabla_\alpha\phi \nabla_\nu\nabla_\beta\phi ,
\end{eqnarray}
where $\hat G$ is the Gauss-Bonnet term, $P^{\mu\nu\alpha\beta}=-1/2 ~\varepsilon^{\alpha\beta\lambda\tau} R_{\lambda\tau\sigma\rho} \varepsilon^{\sigma\rho\mu\nu}$ is the double-dual Riemann tensor, and functions $V_J$, $V_G$, $V_P$, $V_R$ are interaction potentials.

Thus the Covariant Galileons provide an opportunity to extend GR in a way that might explain the smallness of the cosmological constant. One may take a Lagrangian from the Fab Four class and introduce some beyond Fab Four terms. Such a model is going to screen the cosmological constant when Fab Four terms are dominant in the Lagrangian, so the model has a room for a matter-dominated phase of the universe expansion. When beyond Fab Four terms cannot be neglected, the model looses the ability to screen the cosmological constant and the universe enters the late-time acceleration phase. In paper~\cite{Starobinsky:2016kua} the simplest example of such a model is presented. The model includes the John Lagrangian with a constant potential $V_J$, the George Lagrangian with a constant potential $V_G=M_\text{Pl}^2$ and the standard kinetic term for a scalar field (that belongs to $\mathcal{L}_2$ of the Covariant Galileons). Following the aforementioned logic, the model provides a uniform description of inflation, matter-dominated phase of the expansion, and the late-time acceleration of the universe. 

The approach appears to be very fruitful, but it can be broken down at the quantum level, as was pointed in the original paper \cite{Charmousis:2011bf}. In realms of a flat spacetime the Galileons are protected from quantum corrections~\cite{Luty:2003vm, Hinterbichler:2010xn}, a similar situation might appear in a realm of a curved spacetime. Counter terms associated with one-loop Covariant Galileon corrections to the Fab Four interaction might not lie in the Fab Four class. In such a case quantum corrections might spawn a number of terms that cannot be neglected and are unable to screen the cosmological constant, thus ruining the desirable feature of the Fab Four class. An exapmle of a similar behaviour was recently found in the cubic Covariant Galileons \cite{Saltas:2016awg,Saltas:2016nkg}, within the gauge-inveriant regularization scheme Galileons appears to receive additional higher-derivative terms in one-loop effective lagrangian.

In this paper we present a study of a simple model that contains both Fab Four and Covariant Galileons interactions. We consider the John interaction because of the following. First, as was pointed in ref. \cite{Charmousis:2011bf} the Ringo term unable to screen the cosmological constant, it just do not ruins the screening features of a model. Second, only the John and Paul terms provide an example of a non-minimal kinetic coupling that might posses some interesting properties, but the Paul term appears to demonstrate a pathological behavior in star-like objects~\cite{Appleby:2015ysa, Maselli:2016gxk}. Finally, as we pointed before, the John Lagrangian provides a satisfactory description of the evolution of the universe in a similar model~\cite{Starobinsky:2016kua}, which makes it the most perspective candidate for research. Following paper~\cite{Starobinsky:2016kua}, we include the standard kinetic term for a scalar field in the model. As the additional beyond Fab Four interaction that generates loop corrections we consider a standard $\phi^3$ interaction as it is the most simple scalar field self-interaction. 

This paper is organized as follows. In Section~\ref{mainSection} we formulate the task and provide the solution to it. We show that corrections to the John interaction require counter terms that lies in the Fab Four class up to a surface turm. We discuss our results in Section~\ref{Conclusion}. In Appendices we present the full set of Feynman rules for the model and a detailed discussion of infrared divergences of the model.

\section{Quantum corrections and renormalization}\label{mainSection}

As we mentioned before, we consider the simplest scalar field self-interaction $\phi^3$ along with the John interaction. We also include the standard kinetic term for the scalar field following paper~\cite{Starobinsky:2016kua}. In this way we consider the following set of the Fab Four parameters:
\begin{eqnarray}
  V_R = V_P =0 ~, & V_J = \beta_0 + \beta_1 \phi  ~, & V_G = \cfrac{1}{16\pi G} .
\end{eqnarray}
The case $\beta_1=0$ corresponds to the simplest situation, however, we consider $\beta_1 \not =0$. The cubic scalar field self-interaction results in the interaction of a graviton with three scalar particles (the interaction of a graviton with a part of the matter stress-energy tensor associated with a scalar field potential); the John interaction with $\beta_1 \not =0$ also describes an interaction of a graviton with three scalar particles. Thus, this interactions interfere with each other and might bring new divergences into the model, this is the reason to take the case $\beta_1 \not =0$ into account. So, throughout the paper we consider the following action:
\begin{eqnarray}
  S &=& \int d^4 x \sqrt{-g} ~ \left[ \cfrac{1}{16\pi G} R ~ + \cfrac12 ~\pd_\mu \phi~ \pd^\mu \phi + \cfrac{\lambda}{3!} \phi^3 ~ + V_J (\phi) G^{\mu\nu} ~\pd_\mu \phi \pd_\nu\phi \right].
\end{eqnarray}
We are working within a low-energy regime and can use the Effective Field Theory approach to gravity~\cite{Burgess:2003jk, Calmet:2013hfa, Donoghue:2012zc}. Following the standard linearization procedure within the harmonic gauge $( g^{\mu\nu} \Gamma^\lambda_{\mu\nu} =0)$ one obtains the following effective Lagrangian:
\begin{eqnarray}
  \mathcal{L} & = & -\cfrac12 ~h^{\mu\nu} C_{\mu\nu\alpha\beta} \square h^{\alpha\beta} -\cfrac12 \phi \square \phi +\cfrac{\lambda}{3!} \phi^3 + \cfrac{\gok}{2} \cfrac{\lambda}{3!} \phi^3 h -\cfrac{\gok}{2} \pd_\mu\phi\pd_\nu\phi C^{\mu\nu\alpha\beta}(1 + \beta_0 \square +\phi \beta_1 \square) h_{\alpha\beta} ,
\end{eqnarray}
where $h_{\mu\nu}$ is a small metric perturbation, $\gok^2 = 8\pi G$ is the gravitational coupling and $C_{\mu\nu\alpha\beta}$ is defined as
\begin{eqnarray}
  C_{\mu\nu\alpha\beta} &=& \eta_{\mu\alpha} \eta_{\nu\beta} + \eta_{\mu\beta}\eta_{\nu\alpha}-\eta_{\mu\nu}\eta_{\alpha\beta} .
\end{eqnarray}
For the sack of simplicity we use the following terminology throughout the paper. We call $G^{\mu\nu} \pd_\mu \phi \pd_\nu \phi$ as the standard John interaction, $G^{\mu\nu} \phi \pd_\mu \phi \pd_\nu \phi$ as the cubic John interaction, and $\phi^3$ as the cubic self-interaction. We present the expressions for the Feynman rules of the model in Figure~\ref{Feynman_rules_figure} in Appendix A.

\begin{figure}
  \centering
  \begin{minipage}{.32\textwidth}
    \includegraphics[width=.7\textwidth]{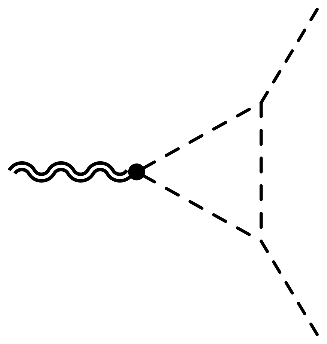}
    \caption{Diagram for the first one-loop amplitude.}
    \label{triangle_diagram_fig}
  \end{minipage}
  \begin{minipage}{.32\textwidth}
    \includegraphics[width=.7\textwidth]{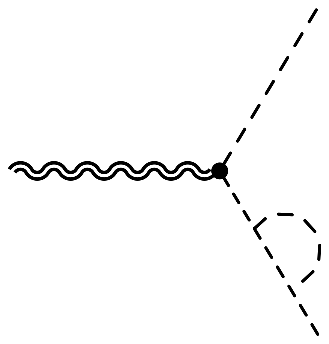}
    \caption{Diagram for the second one-loop amplitude.}
    \label{bubble_diagram_fig}
  \end{minipage}
  \begin{minipage}{.32\textwidth}
    \includegraphics[width=.7\textwidth]{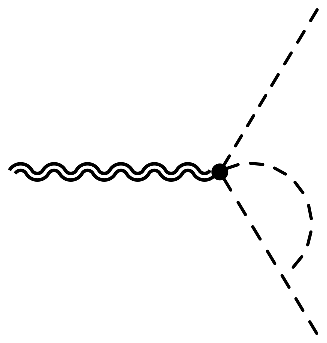}
    \caption{Diagram for the third one-loop amplitude.}
    \label{four_vertex_fig}
  \end{minipage}
\end{figure}

At the first order of the perturbation theory beyond the tree level one should consider only three one-loop amplitudes that affect the standard John interaction, the corresponding diagrams are presented in Figures~\ref{triangle_diagram_fig}, \ref{bubble_diagram_fig}, and \ref{four_vertex_fig}. We present the detailed expressions for amplitudes for Figure~\ref{Loop_amplitudes_fig} in Appendix A, while here we discuss their properties. In the diagrams, the doted vertices mark the interactions from the John class. Amplitudes for Figures~\ref{triangle_diagram_fig} and \ref{bubble_diagram_fig} contain infrared divergences which are canceled out by soft-particle radiation processes. These divergences are generated by the cubic self-interaction, they are not related to the John interaction and can be treated easily. As a matter of fact, these divergences are very similar to infrared divergences in quantum electrodynamics (we provide a more detail discussion in Appendix B). We also would like to underline that there are two types of interactions mixed in the calculation: the standard GR interaction and the Fab Four one. We discuss the corresponding counter terms separately as they play different roles in the task.

The amplitude corresponding to Figure~\ref{bubble_diagram_fig} describes a correction related to the scalar field self-energy and this correction by no means is related to the Fab Four. This correction alone can be excluded from the consideration by a proper (re)definition of the Galileon field asymptotic states. Without the John interaction the expression can be renormalized by the standard kinetic counter term for the scalar field. When the John interaction is taken into account the expression also requires the John-like counter term for renormalization. In such a way, the first amplitude can be renormalized completely within the original Lagrangian.

In the amplitude corresponding to Figure~\ref{triangle_diagram_fig}, only $R_3$ term (see Appendix A for notations) contains ultraviolet divergences that require renormalization. Without the John interaction the amplitude requires only the standard scalar field mass-like counter term to be renormalized, this result should be understand as follows. First, without the John interaction the process contributes to renormalization of the matter stress-energy tensor, so it is not strongly related to the gravitational interaction, rather to the behavior of a massless scalar field. Second, it is known that a quantum massless scalar field requires introduction of a mass scale in realms of a flat spacetime~\cite{Coleman:1973jx}. So we treat the appearance of a mass-like counter term as the standard feature of a massless scalar field that is not related neither with gravity nor with the Fab Four class. We do not discuss this feature of the model below as it is not related with the subject of our research and appears to be an inalienable property of a massless scalar field.

When the John interaction is taken into account the expression corresponding to Figure~\ref{triangle_diagram_fig} also requires the following counter term:
\begin{eqnarray}\label{counter-term_1}
  C_1 = R \phi^2 ,
\end{eqnarray}
where $R$ is the scalar curvature. This counter term belongs to the George class of the Fab Four class. Our model already contains one term from the George class, namely the standard GR Lagrangian which is generated by $V_G=1/16\pi G$. Thus, our result indicates that one has to expand the George potential up to the second order terms:
\begin{eqnarray}
  V_G = \cfrac{1}{16\pi G} + \alpha \phi^2 ,
\end{eqnarray}
where $\alpha$ is a new coupling. The appearance of~\eqref{counter-term_1} does not change desirable features of the model, it rather points to a deep physical connection between the John and George classes. 

The expression corresponding to Figure~\ref{four_vertex_fig} without the John interaction requires only the standard scalar field mass-like counter term. The situation is identical with the previous case and the same logic is applied. When the John interaction enters the consideration, the amplitude requires three more counter terms. The interaction of a graviton with three scalar particles belongs to the John class and has a structure similar to the standard John interaction. However, because of the difference in the structure of interactions correspondent counter terms can be reduced to the standard John (counter) term up to the surface term
\begin{eqnarray}\label{surface_counter-term}
  C_2 = \pd_\mu \left(\sqrt{-g}~ G^{\mu\nu} ~\phi \nabla_\nu \phi \right) ~.
\end{eqnarray}
In the contex of our task the term may be neglected, but it may play an important role in a cosmological task. Thus we showed that the Standard John interaction appears to be renormalizable up to the surface turm.

\section{Discussion and conclusion}\label{Conclusion}

In the paper we showed that the simplest quantum correction to the John Lagrangian of the Fab Four class appears to be renormolizable. We considered the John Lagrangian as it provides the simplest example of scalar-tensor gravity with second order field equations and with the ability to screen the cosmological constant. The model is fruitful as it provides a uniform description of the evolution of the universe from inflation till the late-time accelerated expansion~\cite{Starobinsky:2016kua}. The model has this feature because of the ability to screen the cosmological constant at a certain epoch of the evolution providing a space to the matter-dominated phase of the expansion. In such a way the ability to screen the cosmological constant becomes crucial for the cosmological applications of the model.

At the same time, the model is not protected from quantum corrections and they might ruin that desirable feature. We studied the simplest case of such quantum corrections coming from the standard cubic scalar field self-interaction. Our results show that such corrections require introduction of one counter term~\eqref{counter-term_1} that does not belong to the initial Lagrangian.

Counter term~\eqref{counter-term_1} belongs to the Fab Four class, thus it cannot ruin the desirable property of the model. Moreover, \eqref{counter-term_1} belong to the George class that is already present in the model, thus it only indicates the necessity to take into account higher order terms of the George potential expansion into account:
\begin{eqnarray}
  V_G = \cfrac{1}{16\pi G} + \alpha \phi^2 .
\end{eqnarray}
This result shows that the John and George classes have a non-trivial physical connection and the standard John interaction requires the presence of the George interaction for renormalization.

Because of the massless scalar field infrared issues \cite{Coleman:1973jx} the model requires the standard mass-like counter term for renormalization. However, the issue is related with the properties of the scalar field and by no means related with the gravity, therefore we did not discuss this isse of the model. The rest counter terms can be reduced to the standard John Lagrangian up to the surface term \eqref{surface_counter-term}. 

Summarizing all of the above, we demonstrated that the John and George classes of the Fab Four have a physical connection thought the standard scalar field self interaction. A proper choice of the George poterntial may lead to the renormalizability of the model at the level of one-loop amplitudes.

\section{Acknowledgements}
This work was supported by Russian Foundation for Basic Research via grant RFBR \No 16-02-00682. The authors are also grateful to Antonio Padilla for a very useful comment on surface turms.

\section*{Appendix A. Feynman rules and one-loop amplitudes}

\begin{figure}[h]
  \includegraphics[width=\textwidth]{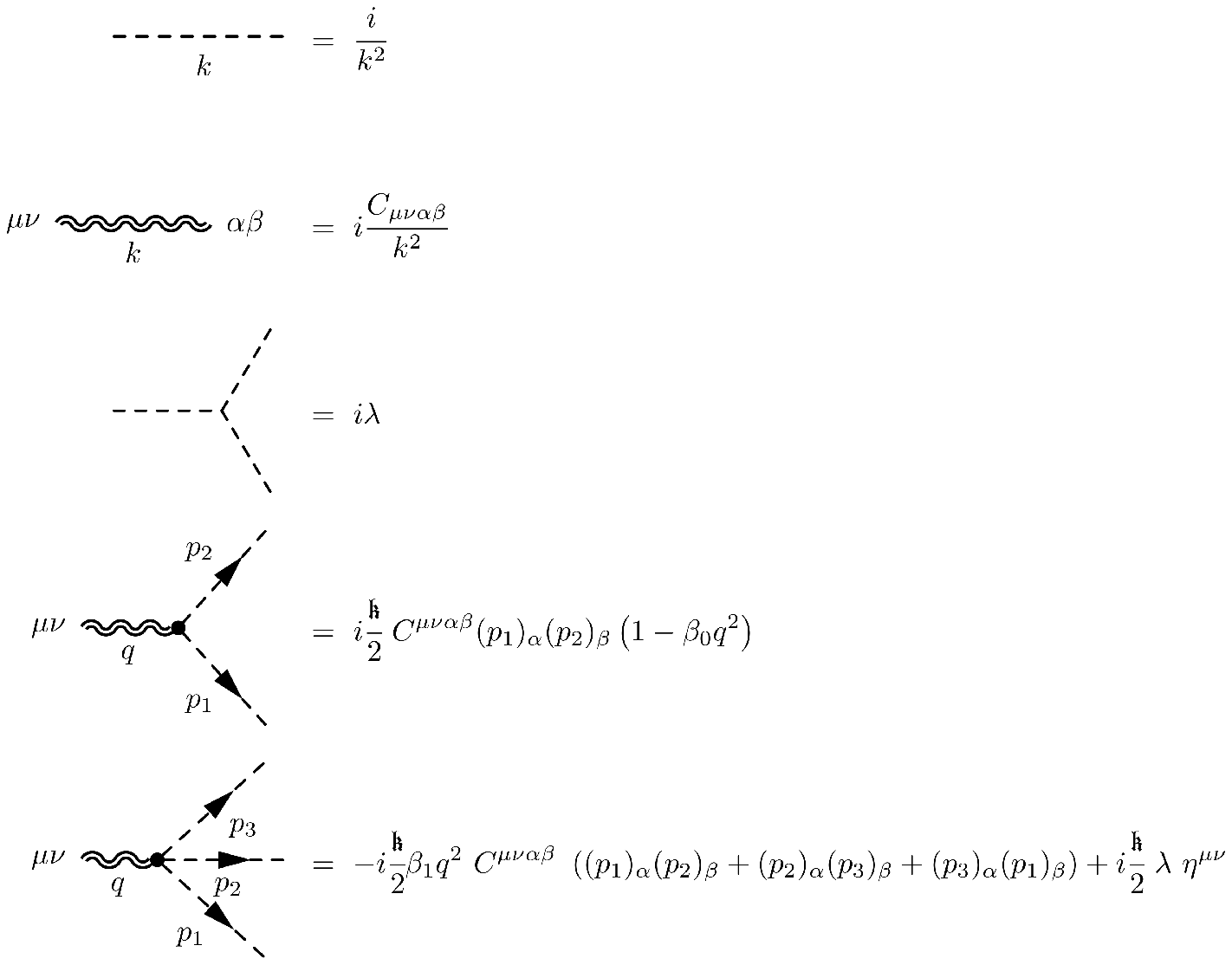}
  \caption{Feynman rules for the model.}
  \label{Feynman_rules_figure}
\end{figure}

\begin{figure}[h]
  \includegraphics[width=\textwidth]{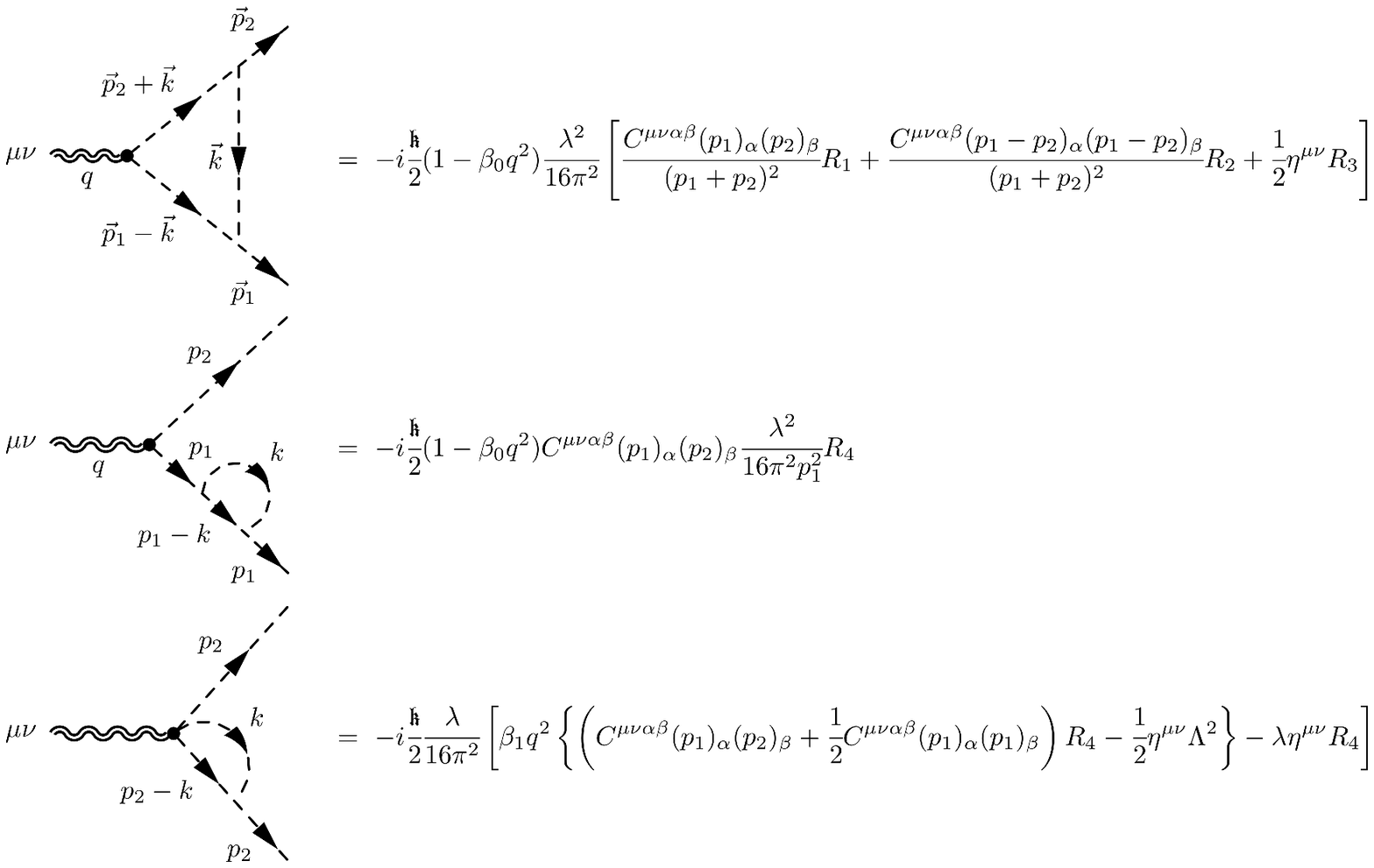}
  \caption{Expressions for one-loop amplitudes.}
  \label{Loop_amplitudes_fig}
\end{figure}

Quantities used in the article have the following mass dimensions: $[\phi]=[\lambda]=1$, $[\gok]=-1$, $[\beta_0]= -2$, $[\beta_1]=-3$. Feynman rules are given in Figure~\ref{Feynman_rules_figure}. In all diagrams momenta are pointed right and the following definitions for $C$ and $D$-symbols are used~\cite{Han:1998sg}:
\begin{eqnarray}
  C_{\mu\nu\alpha\beta} &=& \eta_{\mu\alpha} \eta_{\nu\beta} + \eta_{\mu\beta} \eta_{\nu\alpha} - \eta_{\mu\nu} \eta_{\alpha\beta} ,\\
  D_{\mu\nu\alpha\beta} (k_1,k_2) &=& \eta_{\mu\nu} k_{1\alpha} k_{2\beta} -\left[ \eta_{\mu\alpha} k_{1\nu} k_{2\beta} +\eta_{\mu\beta} k_{1\alpha} k_{2\nu} -\eta_{\alpha\beta} k_{1\mu} k_{2\nu} + (\mu \leftrightarrow \nu) \right] .
\end{eqnarray}
Expressions for one-loop amplitudes discussed in the paper are presented in Figure~\ref{Loop_amplitudes_fig} and the following set of definitions is used:
\begin{eqnarray}
  \int\cfrac{d^4 k}{(2\pi)^4} \cfrac{1}{k^2 (k-p_1)^2 (k+p_2)^2} = \cfrac{i}{16\pi^2} \cfrac{1}{(p_1+p_2)^2} R_1 ~, & & R_1 = \cfrac{\Gamma(1-\varepsilon) \Gamma^2 (\varepsilon)}{2\varepsilon \Gamma(2\varepsilon)} \left( - \cfrac{1}{4\pi} \cfrac{(p_1+p_2)^2}{\mu^2}\right)^\varepsilon ,\\
  \int\cfrac{d^4 k}{(2\pi)^4} \cfrac{k_\mu}{k^2 (k-p_1)^2 (k+p_2)^2} = \cfrac{i}{16\pi^2} \cfrac{(p_1-p_2)_\mu}{(p_1+p_2)^2} R_2~,& & R_2= \cfrac{\Gamma(1-\varepsilon) \Gamma(\varepsilon) \Gamma(1+\varepsilon)}{(1+2\varepsilon) \Gamma(1+2\varepsilon)} \left(-\cfrac{1}{4\pi} \cfrac{(p_1+p_2)^2}{\mu^2}\right)^\varepsilon , \\
  \int\cfrac{d^4 k}{(2\pi)^4} \cfrac{C_{\mu\nu\alpha\beta} k^\alpha k^\beta}{k^2 (k-p_1)^2 (k+p_2)^2} = -\cfrac12 \eta_{\mu\nu} \cfrac{i}{16\pi^2} R_3~, & &R_3 = \cfrac{\Gamma(\varepsilon) \Gamma^2(1-\varepsilon)}{\Gamma(2-\varepsilon)} \left(-\cfrac{1}{4\pi} \cfrac{(p_1+p_2)^2}{\mu^2}\right)^\varepsilon , \\
  \int\cfrac{d^4 k}{(2\pi)^4} \cfrac{1}{k^2 (k+p)^2} = \cfrac{i}{16\pi^2} R_4~, & & R_4 = \cfrac{1}{\varepsilon \Gamma(2+\varepsilon)} \left( \cfrac{1}{4\pi} \cfrac{\Lambda^2}{\mu^2}\right)^\varepsilon ,\\
  \int\cfrac{d^4 k}{(2\pi)^4} \cfrac{k_\mu}{k^2 (k+p)^2} = -\cfrac12 p_\mu \cfrac{i}{16\pi^2} R_4 . & & 
\end{eqnarray}
Factors $R_1$ and $R_2$ contain infrared-divergent parts of the amplitudes, while $R_3$ and $R_4$ contain ultraviolet ones. Parameter $\Lambda$ is used as the ultraviolet cut-off.

\section*{Appendix B. Infrared stability}\label{IR_problem}

The amplitude corresponding to Figure~\ref{triangle_diagram_fig} contains the following infrared-divergent part:
\begin{eqnarray}
-\cfrac{\gok}{2} (1-\beta_0 q^2) \lambda^2 C^{\mu\nu\alpha\beta} (p_1)_\alpha (p_2)_\beta \int\cfrac{d^4 k}{(2\pi)^4} \cfrac{1}{k^2 (k-p_1)^2 (k-p_2)^2} .
\end{eqnarray}
In this expression integration over $k^0$ should be performed:
\begin{eqnarray}\label{I_1}
  \int\cfrac{d^4 k}{(2\pi)^4} \cfrac{1}{k^2 (k-p_1)^2 (k-p_2)^2} &= & \cfrac{i}{8} \left. \int\cfrac{d^3 \vec k}{(2\pi)^3} \cfrac{1}{k^0} \cfrac{1}{k \cdot p_1 ~ k \cdot p_2} \right|_{k^2 =0} - \nonumber \\
  & & \left. \cfrac{i}{8} \int\cfrac{d^3 \vec k}{(2\pi)^3} \cfrac{1}{k^0} \left\lbrace \cfrac{1}{(k \cdot p_1) (k \cdot (p_1 -p_2) + p_1 \cdot p_2)} + [p_1 \leftrightarrow p_2 ] \right\rbrace \right|_{k^2 =0} .
\end{eqnarray}
In \eqref{I_1} only the first term diverges in the infrared sector. Divergent integral is the standard one that appears in soft particle radiation amplitudes, therefore the divergent part of the amplitude is regularized by a soft particle radiation.

The same logic holds for the amplitude corresponding to Figure~\ref{bubble_diagram_fig}. Divergent part of the expression is given by the following:
\begin{eqnarray}
i \cfrac{\gok}{2} (1-\beta_0 q^2) C^{\mu\nu\alpha\beta} (p_1)_\alpha (p_2)_\beta ~i \cfrac{\lambda^2}{p_1^2} \int\cfrac{d^4 k}{(2\pi)^4} \cfrac{1}{k^2 (k-p_1)^2} .
\end{eqnarray}
In a similar way integration over $k^0$ can be performed:
\begin{eqnarray}\label{I_2}
 \cfrac{1}{p_1^2} \int\cfrac{d^4 k}{(2\pi)^4} \cfrac{1}{k^2 (k-p_1)^2} = - \left. \cfrac{i}{4} \int\cfrac{d^3 \vec k}{(2\pi)^3} \cfrac{1}{k^0} \cfrac{1}{(k \cdot p_1)^2} \right|_{k^2 =0} .
\end{eqnarray}
In the full analogy with the previous result we obtain the standard integral for the soft particle radiation amplitude. At the same time, expression~\eqref{I_2} looks as if factor $1/2$ was missing, but this is not the case. All scalar particles in our theory are identical, thus additional permutation of outgoing particles is necessary. As a consequence, the corresponding soft particle radiation amplitude receives the additional factor $2$ and exactly compensates by the infrared divergent term.

%
\bibliographystyle{unsrt}
\bibliography{biblio}

\end{document}